\documentclass[a4paper,twoside]{article}

\usepackage{epsfig}
\usepackage{subcaption}
\usepackage{calc}
\usepackage{amssymb}
\usepackage{amstext}
\usepackage{amsmath}
\usepackage{amsthm}
\usepackage{multicol}
\usepackage{pslatex}

\usepackage{multirow}
\usepackage{algorithm}
\usepackage{algpseudocode}
\usepackage[table]{xcolor}
\usepackage{natbib}
\usepackage{SCITEPRESS}     % Please add other packages that you may need BEFORE the SCITEPRESS.sty package.

\begin{document}

\title{Device-based Image Matching with Similarity Learning by Convolutional Neural Networks that Exploit the Underlying Camera Sensor Pattern Noise}

\author{\authorname{Guru Swaroop Bennabhaktula\sup{1}\orcidAuthor{0000-0002-8434-9271}, Enrique Alegre\sup{2}\orcidAuthor{0000-0003-2081-774X}, Dimka Karastoyanova\sup{1}\orcidAuthor{0000-0002-8827-2590}\\ and George Azzopardi\sup{1}\orcidAuthor{0000-0001-6552-2596}}
\affiliation{\sup{1}Bernoulli Institute for Mathematics, Computer Science and Artificial Intelligence, \\University of Groningen, The Netherlands}
\affiliation{\sup{2} Group for Vision and Intelligent Systems, Universidad de Le\'on, Spain}
\email{g.s.bennabhaktula@rug.nl, enrique.alegre@unileon.es, \{d.karastoyanova, g.azzopardi\}@rug.nl}
}

%  word count 70-200.
\abstract{One of the challenging problems in digital image forensics is the capability to identify images that are captured by the same camera device. This knowledge can help forensic experts in gathering intelligence about suspects by analyzing digital images.
In this paper, we propose a two-part network to quantify the likelihood that a given pair of images have the same source camera, and we evaluated it on the benchmark Dresden data set containing 1851 images from 31 different cameras. To the best of our knowledge, we are the first ones addressing the challenge of device-based image matching. Though the proposed approach is not yet forensics ready, our experiments show that this direction is worth pursuing, achieving at this moment 85 percent accuracy. This ongoing work is part of the EU-funded project 4NSEEK concerned with forensics against child sexual abuse.}

% must be Title Cased
\keywords{Source Camera Identification, Image Forensics, Sensor Pattern Noise.}

\onecolumn \maketitle \normalsize \setcounter{footnote}{0} \vfill

\section{\uppercase{Introduction}}
\label{sec:introduction}

% \begin{itemize}
%     \item Start by introduce briefly the 4NSEEK project
%     \item Describe how camera identification can help in the investigation
%     \item Describe the challenges
%     \item describe the aims, objectives and scope of your task
%     \item describe the structure of the document
% \end{itemize}

\noindent With the rapid adoption and consumption of digital content, there have been many instances of illicit material of children being circulated on the Internet, especially in the darknet. Today, law enforcement agencies (LEAs) require forensic tools which can help them to investigate more effectively and efficiently such digital content. The EU-funded 4NSEEK
project
\footnote{https://www.incibe.es/en/european-projects/4nseek}
, to which this work belongs, is aimed to develop a forensic tool by various partners in the industry and academia with the cooperation of police agencies in the European Union. 
The project is focused on fighting against child sexual abuse and the distribution of its contents across the internet. One desired functionality is device-based image matching, that is the determination whether any two or more seized images were captured by the same camera device. Here we report the ongoing work in this direction. 

% \paragraph{} 
Just as the bullet traces in a crime scene become a piece of evidence for a weapon, a digital image can become an evidence for a camera. This is possible when we can extract fingerprints from images that (uniquely) characterize the source camera device. Extraction and identification of these fingerprints become more challenging when the photographs are subject to compression, post-processing, and computational photography, among others. Every processing step that alters the original RAW image, including the operations that are performed on the captured image within the camera, plays a role in altering the fingerprint. Together with the increasing use of image processing tools, the extraction of fingerprints becomes even more challenging.

%from a camera to uniquely identify the source camera. 

% This can also be used to establish the authenticity of the digital photographic evidence presented to the judicial courts.  

% additionally adding the description of noise models

%  describe the aims, objectives and scope of your task
The camera signature is embedded in the captured image in the form of noise and some artefacts. Our goal is to extract these fingerprints from given images and use them to determine whether the concerned images were captured by the same camera device. We would like to bring out a subtle difference between the terms camera model and camera device, with the former referring the type of camera (e.g. Nikon D200) and the latter refers to a specific manufactured device (e.g. Nikon D200 - 1, where the last digit represents the unique identifier for the manufactured Nikon D200 devices). In this work, we address image matching by using signatures of the source camera devices. %Henceforth, every reference to source camera identification refers to device identification. 

%where the latter deals with the images that are captured with the same camera model but they may not be from the same physical device. The former goes one step further, by trying to pick a specific camera device from cameras of the same model. 

More formally, the problem that we address in this work is the following: given a pair of images, how likely are they were both captured using the same camera device. We restrict our analysis and discussions to the publicly available Dresden \citep{gloe2010dresden} image data set. We propose a convolutional neural network (CNN) based architecture which is in line with the design of the CNN proposed by \cite{mayer2018learned} for camera model identification. 
% The background and the details of our approach are described in the following sections. 

% \paragraph{} 
The rest of the paper is organized as follows. We start by presenting an overview of the traditional and state-of-the-art approaches in Section \ref{sec:related_work}. In Section \ref{sec:methodology}, we describe the approach for feature extraction and classification of the proposed source camera identification. Experimental results along with the dataset description are provided in Section \ref{sec:experiments}. We provide a discussion of certain aspects of the proposed work in Section~\ref{sec:discussion} and finally, we draw conclusions in Section~\ref{sec:conclusion}.

\section{\uppercase{Related Work}}
\label{sec:related_work}

% \begin{itemize}
%     \item Start by describing when this problem was addressed first
%     \item what were the traditional approaches at that time?
%     \item what was the impact of deep learning with CNNs to this application? Has the performance improved drastically? What are the remaining problems? i.e. Why is it still worth investigating the problem?
%     \item Describe briefly why you think it is sensible that the pipeline you are adopting from another application is sensible for this problem too.
% \end{itemize}

\begin{figure}[h]
%   \vspace{-0.2cm}
  \centering
  {\epsfig{file = 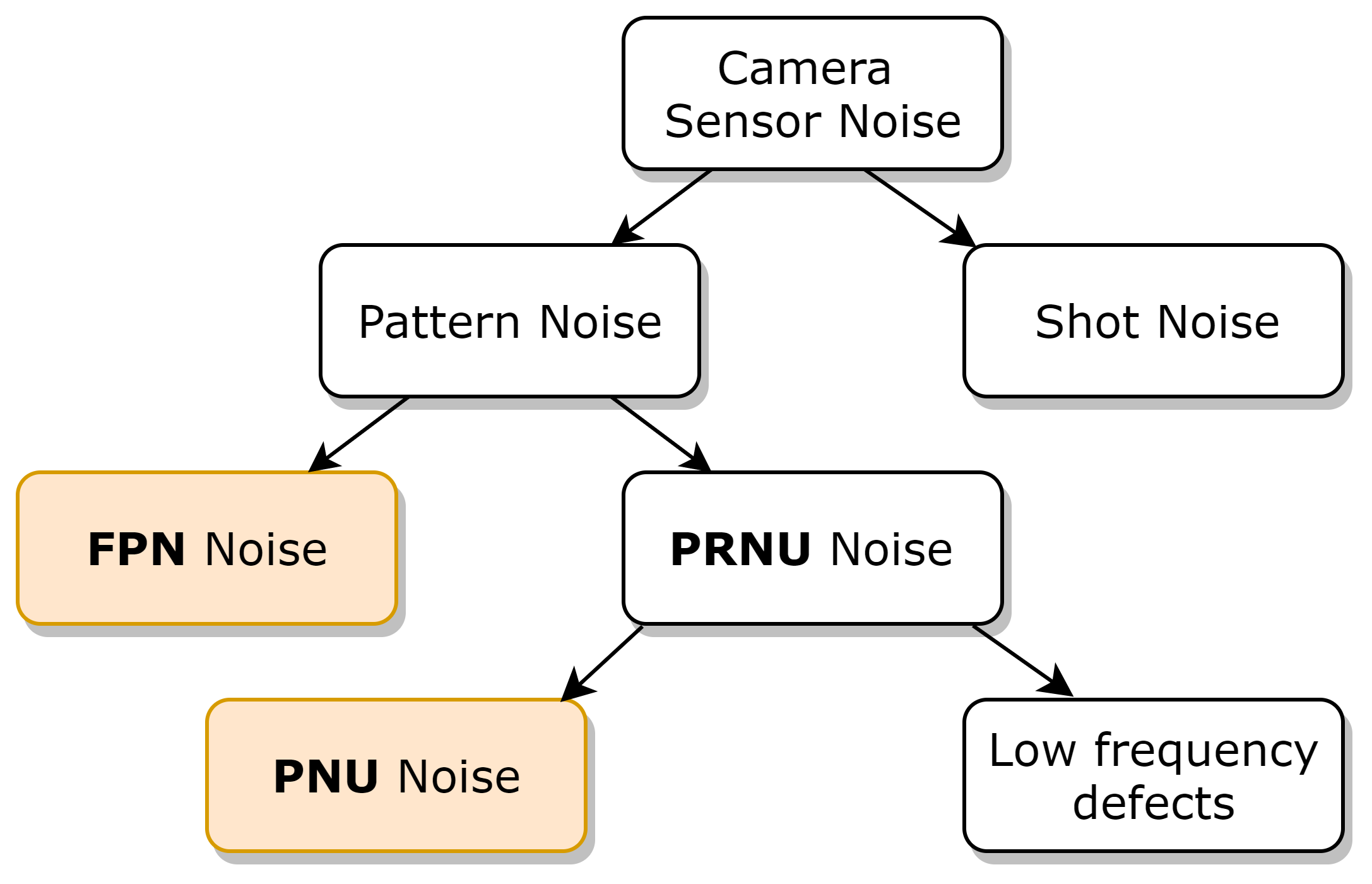, width = 5.5cm}}
  \caption{Topology of digital camera sensor noise. Note that only FPN (Fixed Pattern Noise) and PNU (Pixel Non-uniformity Noise), which are highlighted in yellow, contain the fingerprint that can be used to uniquely identify a sensor.}
  \label{fig:sensor_niose_hierarchy}
%   \vspace{-0.1cm}
\end{figure}

\noindent The camera signature is embedded in the captured image in the form of noise and some artefacts. In Figure \ref{fig:sensor_niose_hierarchy} we illustrate a hierarchical representation of noise classification which we adopt from \cite{lukavs2006digital}. Even when the camera sensor is exposed to a uniformly lit scene the resulting image pixels are not uniform. This non-uniformity is caused due to shot noise and pattern noise. \textit{Shot noise} is a temporal random noise and varies from frame to frame. This component of noise can be suppressed to a large extent by frame averaging. \textit{Pattern noise} is defined as any noise component that survives frame averaging \citep{holst1998ccd}. This stability and uniqueness over time makes pattern noise a candidate for camera signature. 

The two main components of pattern noise are FPN (fixed pattern noise) and PRNU (photo response non-uniform noise). The FPN is an additive noise which is a consequence of dark currents \citep{holst1998ccd}.
Dark currents are responsible for pixel-to-pixel differences when the sensor is not exposed to any light. Some modern digital cameras do offer long exposure noise reduction that automatically subtracts a dark frame from the captured image. This helps in removing the FPN artefacts from the captured image. This is, however, not a de-facto standard and is not implemented by all consumer camera manufacturers.

PRNU is further classified into PNU (pixel non-uniformity noise) and noise caused by low frequency defects. PNU noise is mainly caused due to imperfections and defects introduced into the sensor during the semiconductor wafer fabrication process. This in-homogeneity results in different sensitivity of pixels to light. The nature of PNU is such that even the sensors that are fabricated from the same wafer exhibit different PNU patterns. As mentioned by \cite{lukavs2006digital}, light refraction on dust particles, optical surfaces, and zoom settings also contribute to PRNU noise. These \textit{low frequency} components are not characteristic of the sensor, hence they should be discarded when capturing the noise profile for a sensor from its image.

% \vfill
% The heading of a subsection title must be with initial letters capitalized (titlecased)
\subsection{Traditional Approaches}

\noindent To the best of our knowledge, one of the earliest published works in camera detection was done by \citet{geradts2001methods}. The authors showed that every CCD (charge coupled device) sensor exhibits few random pixels which are defective. These pixels can be identified under controlled temperatures. Repeated experiments showed that the location of such defective pixels always remain the same. The authors built a probabilistic model based on the location of defective pixels. The detection of such pixels is then left to visual inspection. \citet{kharrazi2004blind} proposed 34 handcrafted features combined with an SVM classifier \citep{chang2011libsvm} to distinguish between images taken by Nikon E-2100, Sony DSC-P51, and Canon (S100, S110, S200) cameras. The authors extracted these features from both spatial and wavelet domains and carried out their experiments on a proprietary data set.

\citet{kurosawa1999ccd} were the first to consider FPN for source sensor identification. They established that this type of noise exhibits itself in images and is unique for each camera. The authors observed that the power of FPN is much less than the random noise. Hence, in order to suppress random noise and highlight FPN, they averaged 100 dark frames, which were captured by covering the camera lens. They performed experiments on nine different cameras, eight cameras of which exhibited FPN while the CCD-TRV90 Sony camera did not. \citet{lukavs2006digital} have extended on this work by factoring in PRNU noise in addition to the FPN. For each camera under investigation the authors generated a reference pattern noise, which serves as a unique identification fingerprint for the camera. The reference pattern is generated by averaging the noise obtained from multiple images using a denoising filter. The novelty of that approach is the generation of a camera signature without having access to the camera. Finally, the correlation was used to establish the similarity between the query and reference patterns. 

%\paragraph{}
\citet{li2010source} studied the noise patterns and observed that the scene details have stronger signal components while the true camera noise has weaker signals. Hence, the stronger noise signal components in the residual image should be less trustworthy. Based on this observation, an enhanced noise fingerprint is extracted by assigning less significant weights to strong components of the noise signal.

A variety of techniques have been proposed which account for CFA (color filter arrays) demosaicing artefacts. These methods identify the source camera of an image based on the traces left behind by the proprietary interpolation algorithm used for each digital camera. Notable among these works include those by \citet{bayram2005source,swaminathan2007nonintrusive}, and more recent one by \citet{chen2015camera}. 

\subsection{Approaches based on Deep Learning}
In the last few years, deep learning based approaches have also been applied in the field of image forensics. Several CNN-based systems have been proposed to detect traces of image inpainting \citep{zhu2018deep}, effects of image resizing and compression \citep{bayar2017robustness}, and median filtering detection \citep{chen2015median}, among other image forensic tasks. 

Researchers have additionally proposed to apply CNNs for the identification of source cameras of given images \citep{tuama2016camera,bondi2016first}. 
Most of the deep learning algorithms follow an approach of extracting noise patterns by suppressing the scene content. Interestingly, the first deep learning architectures for image denoising are inspired by the work in steganalysis \citep{qian2015deep}. This is a technique that adds a first layer with a high pass filter which could either be fixed or trainable \citep{bayar2016deep}. \citet{zhang2017beyond} were the first ones to successfully do residual learning by a deep architecture. Residual learning is useful for camera sensor identification because the camera signature is often embedded in the residual images, which are obtained by subtracting the scene content from an image. The authors, proposed a deep CNN model that was able to handle unknown levels of additive white Gaussian noise (AWGN). The CNN model was effective in several image denoising tasks, as opposed to traditional model-based designs \citep{kharrazi2004blind}, which focused on detecting specific forensic traces.
 
The drawbacks of many of the proposed approaches are that they target specific types of forensic traces. For example, researchers have proposed methods that exclusively target CFA interpolation artefacts, chromatic aberration, assume a fixed level of Gaussian noise, and more. This is not an ideal assumption when developing real world applications for forensic investigators. Here, we work with an open set of forensic traces.

The works which are very close to the ideas we propose are those by \citet{cozzolino2019noiseprint} and \citet{mayer2018learned}. Both approaches follow an open set of camera models. Many approaches that rely only on a closed set of camera models rely on prior knowledge from the source camera models. It looks almost impossible to use all existing camera models for training such models, and moreover, the scalability of such systems could be a challenge.

\citet{cozzolino2019noiseprint} designed a CNN which extracts a camera model fingerprint (as an image residue) known as the \textit{noiseprint}. The authors use the CNN architecture proposed by \citet{qian2015deep} and trained it in a Siamese configuration to highlight the camera-model artefacts. Their work primarily focused on the extraction of noiseprint for camera models and on detecting image forgeries. 

The CNN architecture that we adopt in our work is inspired by the work of \citet{mayer2019forensic}. The authors have proposed a system called forensic similarity which determines if two image patches contain the same forensic traces or not. They proposed a two-part network. The first one is a feature extractor and the second part is a similarity network, which determines if two features come from the same source camera model. Patch-based systems do not account for the spatial locality. Therefore, instead of relying only on the patches our proposed system takes the whole image for feature extraction. By considering the whole image the network has the possibility to learn the spatial locality in addition to the sensor pattern noise.

% this also helps in extracting features which account for correlation between patches.  

\section{\uppercase{Proposed approach}}
\label{sec:methodology}
% \begin{itemize}
%     \item Give a highlevel diagram of the architecture of your approach. You already do that
%     \item Try to be more technical here on convincing the reader why the selected pipeline is sensible.
    
% \end{itemize}

\begin{figure}[H]
%   \vspace{-0.2cm}
  \centering
   {\epsfig{file = 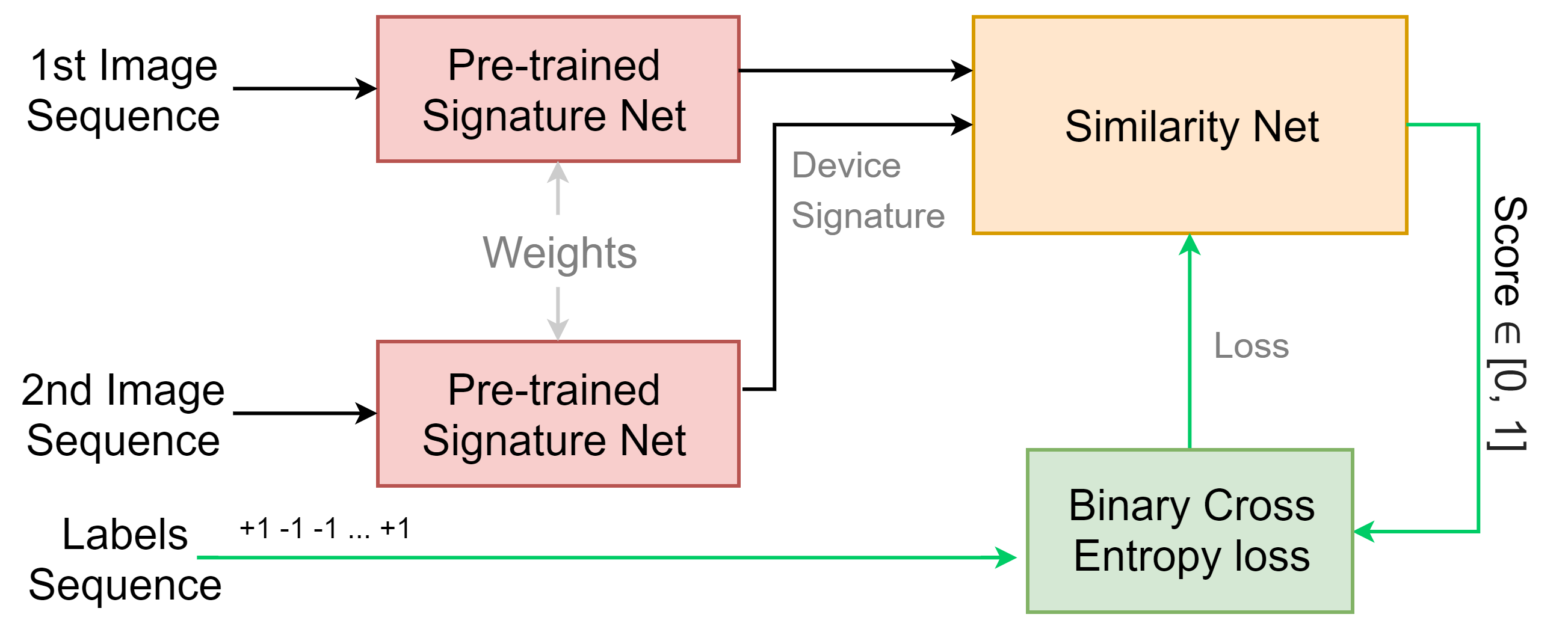, width = 7.5cm}}
  \caption{Proposed workflow.}
  \label{fig:high_level_workflow}
%   \vspace{-0.1cm}
\end{figure}

\noindent The proposed method compares two input images and generates a score indicating the similarity between the source camera devices that took the concerned images. In Figure~\ref{fig:high_level_workflow} we depict the high level workflow of the proposed method. The approach is divided into two phases. In the first phase, we train a CNN called henceforth as \textit{signature network}, responsible for extracting the camera signature from an image. The second stage involves computing the similarity between two image signatures. The similarity function is formulated by training a neural network, which we call \textit{similarity network}. 

A two-phase learning approach gives us the ability to independently fine tune signature extraction and similarity comparison. The training of the networks does not need the availability of ground truth noise residuals. It, therefore, allows us to have a more practical approach, as forensic investigators will not have access to the noise residuals for learning the camera signatures. 

\subsection{Learning Phase I}
The first phase in this approach begins with the training of a signature network, which is defined as follows. 

Let the space of all RGB images be denoted by $\mathbb{I}$. The signature network is trained on a subset of images from $\mathbb{I}$. The trained network is then truncated at a features extraction layer (Layer \# 5, labeled Dense signature in Table \ref{tab:Signature Net architecture}), which we denote by $f_{sig}$. It is a feed-forward neural network function $f_{sig} : \mathbb{I}\times\mathbb{I} \to \mathbb{S}$, where $\mathbb{S}$ is a space of all signatures. We define the signature extraction operation, as follows:  
\begin{equation}
    S = f_{sig}(I)
\end{equation}
$\forall ~I \in \mathbb{I}$, where $S \in \mathbb{S}$.

The signature network consists of four convolutional layers followed by two fully connected layers. A summary of all these layers is shown in Table~\ref{tab:Signature Net architecture}. 
% This is a sequential feed-forward network. 
Note that the number of devices in the final fully connected layer represents the number of camera models present in the training set. The variable $f_{sig}$ represents the trained network truncated at block 5 (see Table \ref{tab:Signature Net architecture}). This gives us a signature of 1024 elements in size. 

\definecolor{maroon}{cmyk}{0,0.87,0.68,0.32}
\begin{table}[t]
\footnotesize
\caption{The proposed CNN architecture of the signature network. It consisting of 4 blocks of convolutional layers and 2 blocks of fully connected dense layers. The highlighted row indicates the layer at which we truncate the network and use the resulting 1024-element feature vector as signature.}
\centering
\begin{tabular}{c | l c c | c} 
  \textbf{\#} & \textbf{Layers} & \textbf{Activation} & \textbf{Dims} & \textbf{Repeat}\\ 
 \hline\hline
 \multirow{3}{*}{1} & Conv 2d & -- & $96\times7\times7$ & \multirow{3}{*}{$\times1$}\\ 
 & Batchnorm & $\tanh$ & -- &\\
 & Max pool & -- & $3\times3$ &\\
 \hline
 & Conv 2d & -- & $64\times5\times5$ & \multirow{3}{*}{$\times2$}\\ 
 2,3 & Batchnorm & $\tanh$ & -- &\\
 & Max pool & -- & $3\times3$ &\\
 \hline
 \multirow{3}{*}{4} & Conv 2d & -- & $128\times1\times1$ & \multirow{3}{*}{$\times1$}\\ 
 & Batchnorm & $\tanh$ & -- &\\
 & Max pool & -- & $3\times3$ &\\
 \hline
 \rowcolor{maroon!10}
 5 & Signature &  tanh &  1024 & $\times1$\\
 \hline
 \multirow{2}{*}{6} & Dense & tanh & 200 & \multirow{2}{*}{$\times1$}\\
 & Dense & softmax & \# devices & \\[0.5ex]
%  \hline 
\end{tabular}
\label{tab:Signature Net architecture}
\end{table}

\subsection{Learning Phase II}
The goal of the second phase is to map the signatures of pairs of images to a similarity score that gives an indication of whether the input pair comes from the same or different source. To this extent, we train a neural network in a Siamese fashion that determines the similarity between a pair of signatures extracted using the signature network. Let $S_1$ and $S_2$ be two signatures extracted from the signature network; $S_1 = f_{sim}(I_1)$ and $S_2 = f_{sim}(I_2)$. The labeled data for training the similarity network is then generated according to the following condition:
\begin{equation}
     S_{label}(S_1, S_2) =
        \begin{cases}
            \multirow{2}{*}{1,} & \text{If $I_1$ and $I_2$ come from}\\
                                & \text{the same source camera}\\
            0, & \text{otherwise} \\
        \end{cases}
    \label{eq:similarity_net_labels}
\end{equation}

\begin{figure}[H]
%   \vspace{-0.2cm}
  \centering
   {\epsfig{file = 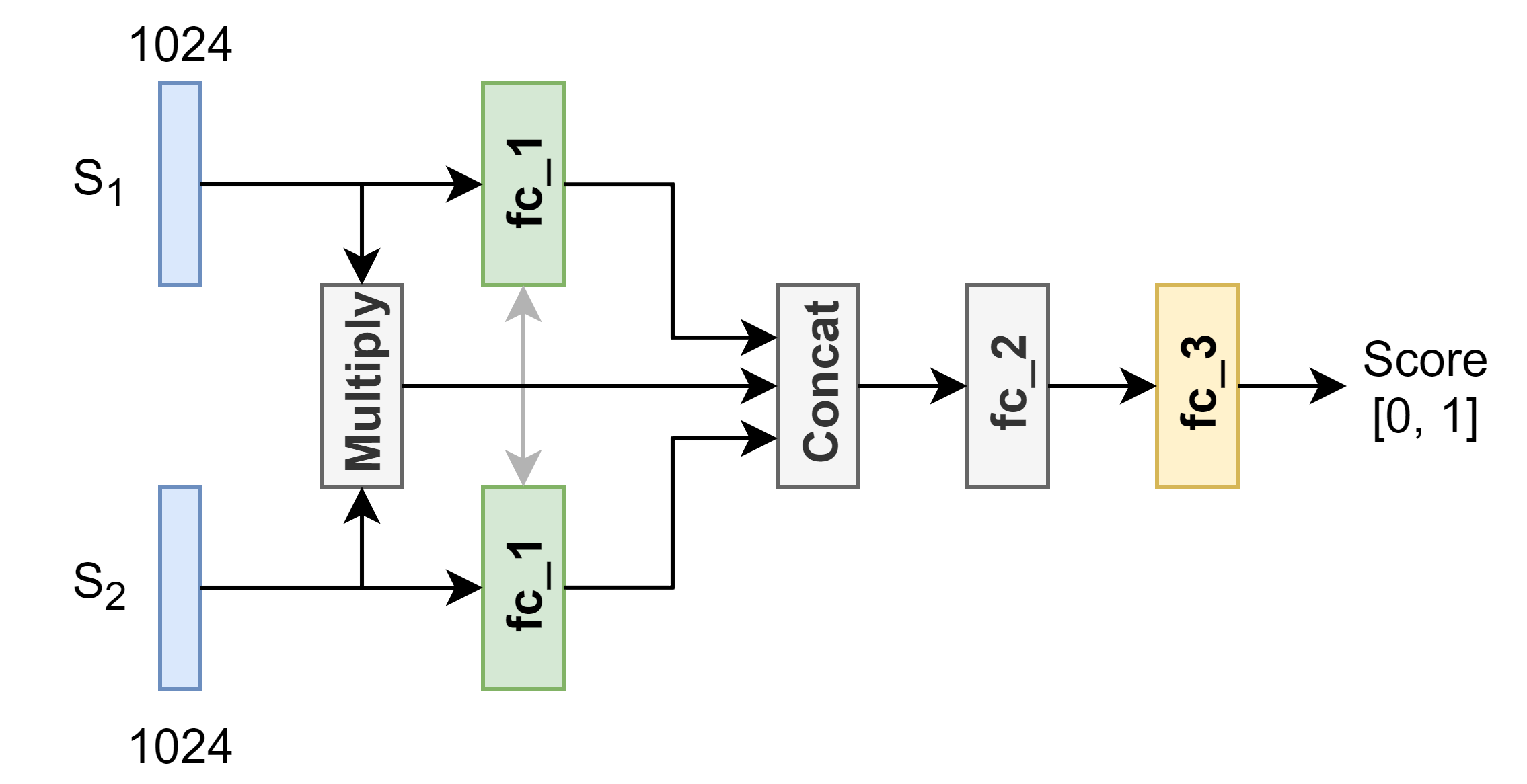, width = 7.5cm}}
  \caption{The proposed neural network architecture of the Similarity Network.}
  \label{fig:similarity_net}
%   \vspace{-0.1cm}
\end{figure}
\paragraph{} 
The similarity network learns the mapping $f_{sim}: \mathbb{S}\times\mathbb{S} \to [0, 1]$, and its architecture is depicted in Figure~\ref{fig:similarity_net}. The first layer is a fully connected dense layer ${fc}\_1$ containing 2048 neurons with ReLU activation, which takes as input the signatures $S_1$ and $S_2$ of a given pair of images. Then, we combine the outputs from the first dense layer along with an element wise multiplication of $S_1$ and $S_2$ into a single vector and feed it to ${fc}\_2$, which is a dense fully connected layer with ReLU activations. This is finally connected to a single neuron with a sigmoid activation.
Once the similarity network is trained, we can use both networks together in a pipeline to determine the similarity for any given pair of input images.
\begin{equation}
    score = f_{sim}(f_{sig}(I_1), f_{sig}(I_2)) 
    \label{eq:score}
\end{equation}
We experimentally determine a threshold $\eta$ for the score given by the network. The pairs of images whose similarity score is above $\eta$ are classified as similar, otherwise as different.

% We have extracted signatures from images that were part of training as well as the validation set used for the signature net.  

\section{\uppercase{Preliminary Experiments and Results}}
\label{sec:experiments}
% \begin{itemize}
%     \item I changed the heading to preliminary experiments and results.
%     \item Describe the experiments that you conducted. 
%     \item Show the results and comment why they are not to your satisfaction
%     \item Create a new section called something like "Further Plans" and describe few ideas of how and why you think they will improve the pipeline.
% \end{itemize}
%We begin by a brief description of the data set, followed by the details of the experiments.
\begin{figure*}[ht]
%   \vspace{-0.2cm}
  \centering
%   {\epsfig{file = sim_matrix_jpeg_test.png, width = 7.5cm}}
    {\epsfig{file = 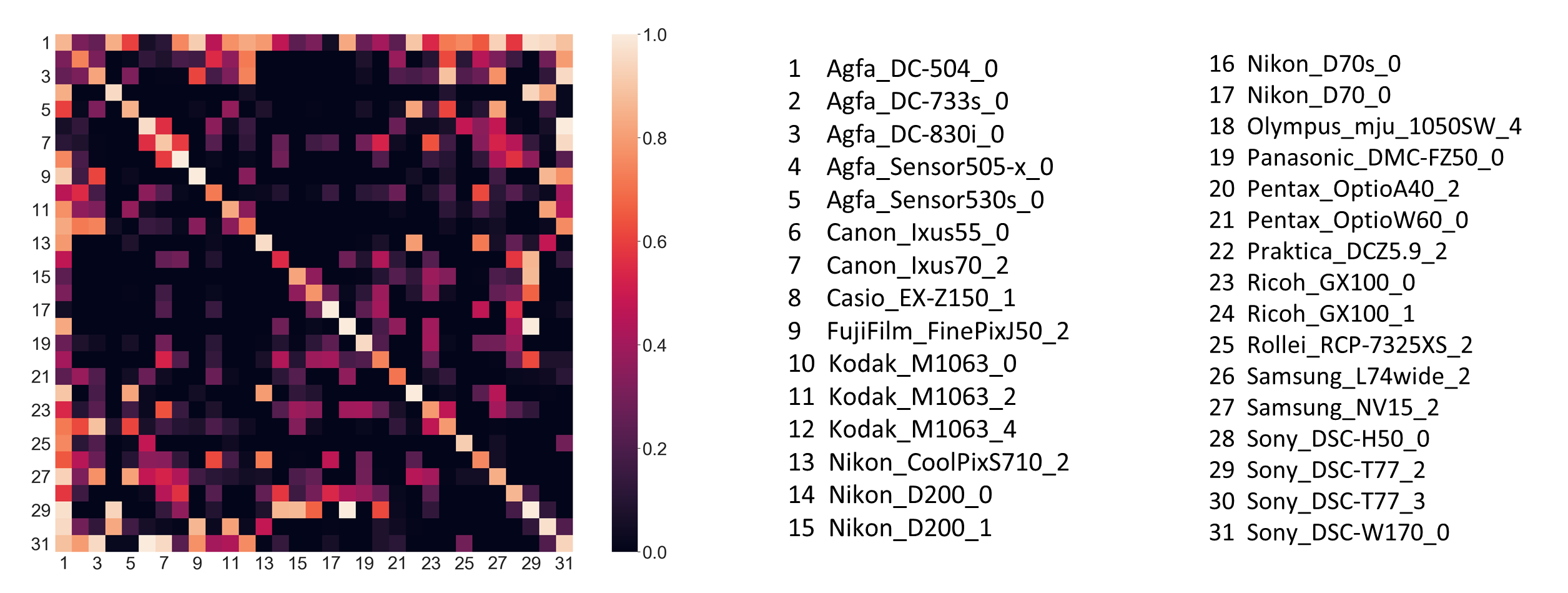, width = 15cm}}
  \caption{Similarity matrix for the 31 camera devices in the test set. A score closer to 1 indicates a high similarity between the images taken from the corresponding pairs of cameras. Similarity values along the diagonal correspond to the similarity between images taken from the same cameras. Ideally, the similarity matrix has ones along diagonal, and zeros elsewhere.}
  \label{fig:sim_matrix}
%   \vspace{-0.1cm}
\end{figure*}

\subsection{Dataset}
\noindent We used the publicly available Dresden dataset \citep{gloe2010dresden} in our experiments for image matching based on source camera identification. It consists of images from various indoor and outdoor scenes acquired under controlled conditions. 

%The images were captured by 73 different devices that come from 25 camera models consisting of 16187 images.

Many camera model identification approaches have been presented but due to a lack of benchmark datasets, it is often hard to directly compare the performance of different methods. The Dresden dataset was made available in 2010 and since then it has seen widespread use in image forensics that also go beyond source camera identification. The Dresden dataset comes with three subsets of data, one of which is called JPEG, which was intended for the study of model specific JPEG compression algorithms. The JPEG set consists of 1851 images taken by 34 different camera devices that belong to 25 camera models. We discard the three camera devices (FujiFilm\_FinePixJ50\_0, Ricoh\_GX100\_3, Sony\_DSC-T77\_1) that contain only one image each and work with the remaining 31 devices. Though the content is limited to two indoor scenes, it is of interest to understand the source camera device identification in the presence of JPEG compression artefacts. The other two subsets, which consist of dark frames and natural images, were not considered in our study.

\subsection{Experiments}
In a random stratified manner, we used 70\% of the data for training and validation. The remaining 30\% of data was left for our tests. 

In the first training phase the signature network was trained for 15 epochs, with categorical cross entropy as the loss function and stochastic gradient descent (SGD) as the optimizer. For the optimization task, we set the learning rate to 0.001, the momentum to 0.95, and the decay to 0.0005. 
Convergence was reached after the ${\text{5}}^{\text{th}}$ epoch where the validation loss started to fluctuate while the training loss remained roughly the same, and we fixed the network with weights obtained at the end of the ${\text{5}}^{\text{th}}$ epoch. The training was done on an NVIDIA RTX 2070 GPU. All the extracted signatures were stored in a database, which provided easy access in the second half of the experiments.

In the second phase, where we train the similarity network, we generated labeled pairs of signatures according to Equation \ref{eq:similarity_net_labels}. All the 1294 (70\% of the full data set of 1851 images) training and validation images used for learning the signature network generated $^{1294}\rm C_2$ pairs of labeled signatures data. The similarity network was trained in a Siamese fashion using binary cross entropy as the loss function along with an SGD optimizer. The network was trained for 30 epochs, with a learning rate of 0.005 and a decay factor of 0.5 for every 3 epochs.

For the systematic evaluation of the trained network, we performed a series of experiments. A single experiment involves choosing a pair of camera devices and generating 100 random pairs of images with replacement. Each pair consists of an image from each of the two concerned camera devices. The trained network was used to predict the similarity score for each of the pairs. The similarity score is converted to 1 (similar) or 0 (not similar) based on a threshold which we determined from the evaluation on the validation set. We set the threshold to 0.99 as it provided the maximum F1-score on the validation set. We normalized the resulting 100 scores by averaging them in order to get a value between 0 and 1 for the comparison of images coming from two camera devices.

For evaluation of the network, we considered all possible pairs of the 31 camera devices resulting in a $31 \times 31$ experiments. We used Algorithm~\ref{alg:evaluation} below to generate a similarity matrix of $31 \times 31$ elements, where each element is the normalized similarity score of the corresponding camera devices. Figure~\ref{fig:sim_matrix} shows the resulting similarity matrix of our test data. The overall accuracy is 85\%.

\begin{algorithm}
  \caption{Similarity matrix computation.}
  \label{alg:evaluation}
    %   \begin{algorithmic}[1] [1] is for displaying line numbers
    \begin{algorithmic}
    \Procedure{Similarity Matrix}{}
      \State $\mathbf{C} \gets \{C_1, C_2, ..., C_{\text{N}}\}$ \Comment{N cameras}
      \For{$i\gets 1:\text{N}$ \texttt{and} $j\gets 1:\text{N}$}
        \State \texttt{Randomly sample 100 image pairs}
        \State \texttt{from the subspace $C_i \times C_j$}
        \State
        \State \texttt{Predict the Source Similarity for}
        \State \texttt{the concerned 100 pairs of images}
        \State \texttt{using Equation \ref{eq:score}.}
        \State
        \State \texttt{Compute the accuracy}
      \EndFor
      \State \textbf{return} accuracy for $\text{N} \times \text{N}$ experiments
    \EndProcedure
  \end{algorithmic}
\end{algorithm}

\section{\uppercase{Discussion and Future Work}}
\label{sec:discussion}

\noindent As can be seen from Figure \ref{fig:sim_matrix}, in general, the model is able to detect images coming from the same camera devices. There are, however, some instances where the network gets confused with images coming from the same camera model. This can be seen with camera models (Ricoh\_GX100\_0, Ricoh\_GX100\_1), (Nikon\_D70\_0, Nikon\_D70s\_0). This could be because the same camera models are subject to the same manufacturing process. Thereby, resulting in similar imperfections or artefacts. We need to first investigate the noise differences between the same camera models, before trying to investigate the noise patterns together from all the devices. This approach might give us a better insight into the challenges between the same camera models.

It is also evident that the devices from the brands Agfa and Sony get confused with several other camera devices in our evaluation. We suspect this is due to the presence of a large number of images in the data set coming from Agfa (around 25 percent), which may have caused some bias in the learned networks. We will address this problem by investigating different approaches that deal with unbalanced training sets.

The approach that we propose mimics the practical situation faced by forensic experts, where they only have a collection of images without knowing their actual source. Among others, investigators are interested to determine whether two or more images were taken by the same camera, irrespective of what camera it is. That information can help them identifying the offender or to compile stronger evidence. To the best of our knowledge, this is the first attempt that addresses the problem of device-based image matching. 

\section{\uppercase{Conclusions}}
\label{sec:conclusion}

\noindent From the results we achieved so far we conclude that the proposed approach is promising for matching images based on their underlying sensor pattern noise. We will continue our investigations and aim to improve the method until it is robust enough to be deployed as a forensic tool.

\section*{ACKNOWLEDGEMENTS}
  This research has been funded with support from the European Commission under the 4NSEEK project with Grant Agreement 821966. This publication reflects the views only of the author, and the European Commission cannot be held responsible for any use which may be made of the information contained therein.

\bibliographystyle{apalike}
{\small
\bibliography{example}}

\end{document}